\documentclass[preprint,showpacs,preprintnumbers,amsmath,amssymb]{revtex4}

\usepackage{graphicx}
\usepackage{dcolumn}
\usepackage{bm}

\begin{document}


\title{
Fracture Toughness and Maximum Stress \\  in a Disordered Lattice System 
}

\author{Chiyori Urabe}
\email{chiyori@ams.eng.osaka-u.ac.jp}
\affiliation{%
Graduate School of Engineering, Osaka University, Osaka 565-0871, Japan
}%
\author{Shinji Takesue}%
 \email{takesue@scphys.kyoto-u.ac.jp}
\affiliation{%
Department of Physics, 
Graduate School of Science, Kyoto University, Kyoto 606-8501, Japan
}%

\date{\today}

\begin{abstract}
Fracture in a disordered lattice system is studied.
In our system, particles are initially arranged on the triangular
lattice and each nearest-neighbor pair is connected with a randomly chosen
soft or hard Hookean spring.
Every spring has the common threshold of stress at which it is cut.
We make an initial crack and expand the system perpendicularly to the
crack.  We find that the maximum stress in the stress-strain curve
is larger than those in the systems with soft or hard springs only
(uniform systems).
Energy required to advance fracture is also larger in some
disordered systems, which
indicates that the fracture toughness improves.  The increase of the
energy is caused by the following two factors. One is that the soft spring is
able to hold larger energy than the hard one.
The other is that the number of cut springs increases as 
the fracture surface becomes tortuous in disordered systems.

\end{abstract}


\maketitle

\section{Introduction}

Composite of hard and soft materials sometimes results in increase of
toughness.  An example is double-network (DN) gel that is an 
interpenetrating network of brittle polyelectrolyte gel and  
flexible polymer chains. A DN gel can have much higher toughness than
ordinary gels \cite{Gong2003,Na2004,Tanaka2005,Tanaka2005_2,Tanaka2007}.
The second example is nacre, which has a layered structure of brittle 
ceramic and soft organic matrix. The toughness of nacre is comparable to 
that of some high-technology structural
ceramics\cite{Curey1977,Jackson1988}. 
In addition, the toughness of brittle $\mathrm{Al_2O_3}$ ceramics is
known to be increased through the incorporation of metals like tungsten,
copper or nickel. The bending strength also increases with the content
of these metals\cite{Tuan1990,Sekino1997,Oh1998,Li2004}.

These materials have been studied as follows.
Tanaka\cite{Tanaka2007} argued that DN gel locally softens and the crack 
extends within the softened zone. He estimated effective fracture
energy, whose order of magnitude is consistent with experiments. 
With respect to nacre, it is considered that a weak material 
reduces drastically the stress concentration near the fracture
tips\cite{deGennes2000} and the maximum stress increases with the repeat
period of structure. Okumura examined a double elastic network of hard
and soft materials and derived toughness enhancement
factors\cite{Okumura2004}, where the ratio of the two elastic moduli and
the mesh size as well as the volume fraction of each element are
especially important in controlling toughness. 
For the composites of $\mathrm{Al_2O_3}$, Ashby, Blunt and Bannister
emphasize that the inclusions bridge the crack and are stretched as 
the crack opens, leading to enhancement of toughness\cite{Ashby1989}.
They executed experiments of bonding a wire into a thick-wall glass
capillary and showed that the toughness increases with the diameter of
wires. It has been derived that the toughness is proportional to 
the square root of the product of volume fraction and inclusion size, 
which is confirmed in experiments by Tuan and Brook\cite{Tuan1990}.

Although structure of the composite materials may be important in the above
examples, we here focus on whether toughness can be improved only by mixing
hard and soft materials. 
Specifically, we consider a model system which consists of particles initially on the triangular lattice.
Every nearest-neighbor pair of the particles is connected with a randomly chosen hard or soft spring except those on the initial crack. 
Each spring is assumed to
break if it is subject to stress larger than a threshold value, which is
common to all springs. We numerically investigate behavior of the system
when it is stretched at constant velocity and find that the system
can hold up larger stress than a uniform system with only a kind of
spring can.
Moreover, the maximum stress depends on the fraction of soft springs and 
the ratio of the spring constants, respectively.

There are some related works in the 
literature\cite{Beale1988,Espanol1996,Politi2001}.
In particular, Beale and Srolovitz studied the system where springs are
present with a probability\cite{Beale1988}.
They obtained analytically the distribution function of the initial 
failure stress and verified it using computer simulation.
Espan\~{o}l, Z\'{u}\~{n}iga and Rubio showed numerically that the stress
of the most stretched spring is represented by a function of the
initial notch length\cite{Espanol1996}.  
To our knowledge, however, enhancement of the maximum stress or
toughness in such a disordered system has not been reported to date. 

We organize the paper as follows.  In the next section, our model and 
simulation method are described.  Then we show our numerical results in
Sec.~\ref{results}. 
The last section is devoted to summary and discussion.

\section{Model and Simulation}

\begin{figure}
\begin{center}
\includegraphics[width=6cm]{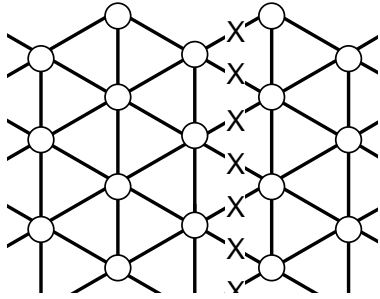}
\caption{\label{figTcut}Upper 20 springs between the $24$th and the $25$th columns are 
initially cut.}
\end{center}
\end{figure}

\begin{figure}
\begin{center}
\includegraphics[width=3cm]{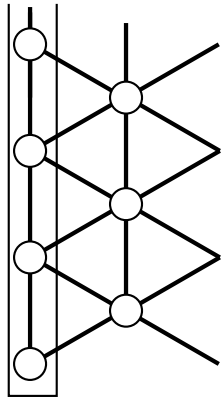}
\caption{\label{figTW} The particles on a wall move as one object.}
\end{center}
\end{figure}

\begin{figure}
\begin{center}
\includegraphics[width=7.5cm]{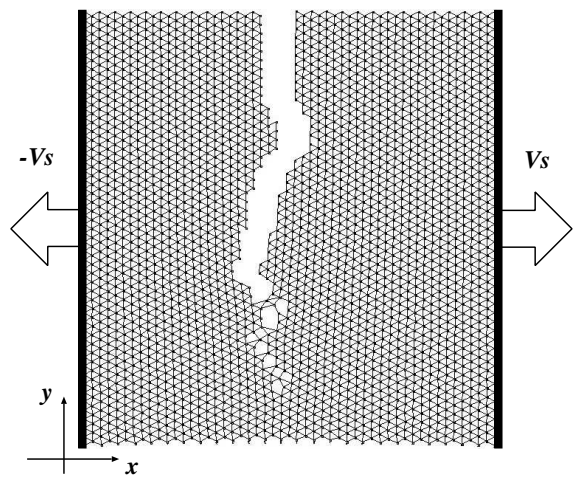}
\caption{\label{fig1} Crack propagates as the walls are moved outward in the horizontal
direction.  The initial crack is introduced in the upper part of the system.}
\end{center}
\end{figure}

Our model system consists of $2525$ particles initially on the triangular 
lattice arranged in $51$ columns each of which has alternatingly $49$ or 
$50$ particles. The particles have the same mass and each pair of
nearest-neighbor particles is connected by a linear spring of natural
length $l_0$, which is equal to the lattice spacing. Spring constant
$k_1$ or $k_2$ ($k_1\ge k_2$) is randomly assigned to each spring
according to a given ratio $1-r:r$. Namely, $r$ denotes the fraction of
soft springs. Each spring is assumed to break if the force on the spring
exceeds a threshold value $\sigma^*$, which is common to all
springs. The broken springs do not generate any elastic force. As shown
in Fig.~\ref{figTcut}, twenty springs from the top between the $24$th and
the $25$th columns of particles are cut in the initial condition. Thus,
the initial crack is about $1/5$ as long as the column length. The
particles on the leftmost (or rightmost) column move as one object, and
we call them walls (Fig.~\ref{figTW}). The walls are moved by an
external force in the outward directions at constant velocity $v_s$ as
shown in Fig.~\ref{fig1}. The walls are assumed to be vertically
incompressible. Moreover, the viscous force with viscosity coefficient
$\eta$ acts on the particles. We assume that $v_s$ is slow and the
viscous force is strong enough to make the dynamics of particles
overdamped. Thus, the motion of the $i$th particle obeys the following
equation,
 \begin{equation}
 \sum_{j}k_{ij}\left(\left|\mathbf{x}_i-\mathbf{x}_j\right|-l_0\right)
 \hat{\mathbf{x}}_{j\to i}+\eta\dot{\mathbf{x}}_i=0,
\label{eq1}
 \end{equation}
where $\mathbf{x}_i$ denotes the position of the $i$th particle, 
$k_{ij}$ is the spring constant of the spring between the $i$th 
and $j$th particles, and $\hat{\mathbf{x}}_{j\to i}$ is the unit vector 
from the $j$th particle to the $i$th one. The summation runs over the
nearest neighbors of the $i$th particle. 

Because length and time can be rescaled with the natural length $l_{0}$ 
and $\eta/k_{2}$, respectively,
we can set $l_{0}=1$  and $\eta/k_2=1$ without loss of generality.
We further assume that $\sigma^{*}=1$ for simplicity.  
Then the remaining system parameters are $k=k_{1}/k_{2}>1$, $r$ and $v_s$.
We call  $k$ the elastic modulus ratio in this paper.

Simulations are basically carried out using the fourth-order 
Runge-Kutta method with the time step $\Delta t=1.0\times 10^{-2}$.  
However, the time step is changed temporally when a spring breaks as 
follows. 
If the computed force of a spring, $\sigma'$, which is below the threshold $\sigma^*$ at time $t_0$ exceeds $\sigma^*$ until $t_0 + \Delta t$,  
we approximate the
change of force between time $t_{0}$ and $t_{0}+\Delta t$ by a linear
function and obtain the new time step $\delta t$ determined by
$\sigma'(t_0+\delta t)=\sigma^*$. That is,
\[
 \delta t = \frac{\sigma^* - \sigma'(t_0)}{\sigma'(t_0+\Delta
 t)-\sigma'(t_{0})}\Delta t.
\]  
After the motion of the particles between $t_{0}$ and 
$t_{0}+\delta t$ are calculated again using the fourth-order 
Runge-Kutta method, the spring is cut.
Then, the next step proceeds with the normal time step $\Delta t$.

We compute stress $\sigma$ and  strain $\varepsilon$ in the horizontal
direction at the walls as 
\begin{eqnarray} 
\sigma & = & 
\frac{1}{H_{0}}\sum_{i=1}^{N_{w}} \sum_{j} 
k_{ij} \left(
\mid\mathbf{x}_{i}-\mathbf{x}_{j}\mid -l_{0}
\right)
\frac{\mid x_{j}-x_{i}\mid}{\mid\mathbf{x}_{j}-\mathbf{x}_{i}\mid},
\label{stress}\\
\varepsilon 
& = & \frac{\mid X_{r}(t)-X_{l}(t)\mid  - W_{0} }{W_{0}}, \label{strain}
\end{eqnarray}
where $N_{w}$ is the number of particles on each wall, $x_{k}$ is the
horizontal coordinate of the $k$th particle, $X_l$ and $X_r$ denote the 
horizontal coordinates of the left and right walls, 
and $H_{0}$ and $W_{0}$ are the height of the walls and the distance 
between the walls at the initial time, respectively.
The summation for $i$ is taken over the particles on the wall, and 
$j$ runs over the nearest-neighbors of each $i$.
Since the values of stress evaluated at the left and right walls are almost 
the same, we do not distinguish them in the following.

\begin{figure}
\begin{center}
\vspace*{1em}
\includegraphics[width=7.5cm]{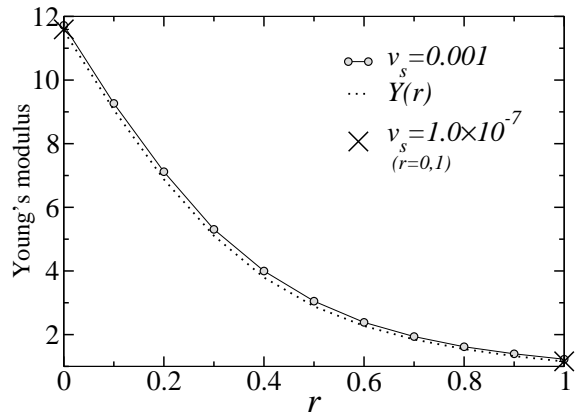}
\caption{\label{YM} 
Young's modulus vs the fraction of soft springs $r$.
Dotted line represents the theoretical estimate by Garboczi and Thorpe.}
\end{center}
\end{figure}

Young's modulus of our system was calculated by Garboczi and 
Thorpe\cite{FengThorpeGarboczi1985,GarbocziThorpe1986_III} by applying the
effective medium theory. It is a kind of mean-field theory which
approximates the random and disordered system with an equivalent uniform
system. The spring constant $k^*$ of the uniform system corresponding to
our system is determined by the relation
\[
 \frac{2(1-r)(k^*-k_1)}{k^*+2k_1}+\frac{2r(k^*-k_2)}{k^*+k_2}=0.
\]
Thus, the Young's modulus $Y$ is obtained as
\begin{eqnarray*}
Y=\frac{1}{\sqrt{3}}\left(-b+\sqrt{b^2+8k}\right)
\end{eqnarray*}
where $b=(k-1)(3r-1)+1$. Because the number of soft springs increases
with $r$, $Y$ is a decreasing function of $r$. Fig.~\ref{YM} shows a
very good agreement between the theory and our numerical results. The
latter is computed from the initial slope of the stress-strain (S-S)
curve. We remark that the system with vertically compressible walls and
no initial cracks is used for the computation. The small deviation is
considered as due to the finiteness of the wall velocity $v_s=10^{-3}$,
since the agreement gets much better when a smaller velocity
$v_s=10^{-7}$ is used. In addition, the Poisson's ratio $\gamma$ is
theoretically computed to be $1/3$, while $\gamma$ lies between $0.29$
and $0.33$ numerically when $v_s=10^{-3}$.  

\section{Results}
\label{results} 

\begin{figure}
\begin{center}
\vspace*{1em}
\includegraphics[width=7.5cm]{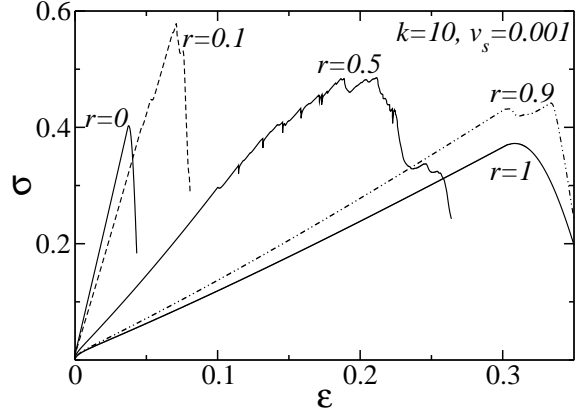}
\caption{\label{SS_H}Stress-strain curve for the systems with various
 $r$ and fixed $k=10$.}
\end{center}
\end{figure}

\begin{figure}
\begin{center}
\vspace*{1em}
\includegraphics[width=7.5cm]{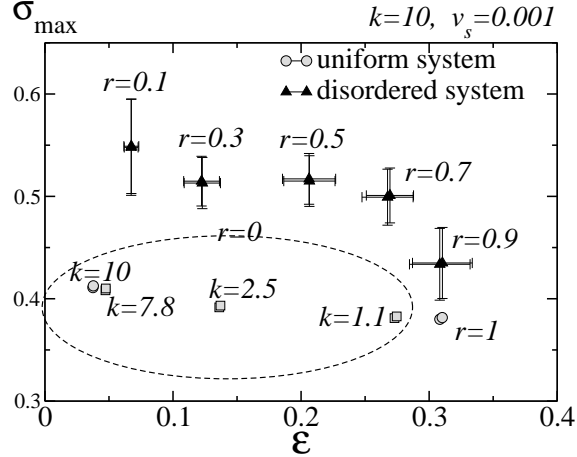}
\caption{\label{topSS_H}The average of the maximum stress for the
 disordered systems with $k=10$ (circles and solid triangles) and those
 for the uniform systems with various $k$ (squares).}
\end{center}
\end{figure}

In Fig.~\ref{SS_H}, we show the S-S curves obtained for the 
models with various $r$ and fixed $k$ at $10$. Stress $\sigma$ initially
increases linearly with strain $\varepsilon$. When the stress reaches a
maximum, the crack begins to propagate. 
The value of the maximum stress varies from
sample to sample in the disordered cases ($0<r<1$). However, for almost
all samples, the maximum stress is larger than those at $r=0$ and $r=1$
despite the fact that the threshold $\sigma^{*}$ is common to the soft
and hard springs. Thus, as shown in Fig.~\ref{topSS_H}, the average
values of the maximum stress exhibit significant enhancement compared
with the uniform cases. We note that this is entirely a nonlinear
effect. Linear properties like Young's modulus and Poisson's ratio are
explained by considering equivalent uniform systems, which show no
increase of the maximum stress. Squares in Fig.~\ref{topSS_H} represent
the maximum stress for uniform models with $r=0$ and $k=1.1, 2.5, 7.8$,
whose Young's modulus correspond to $k=10$ and $r=0.9,0.5, 0.1$,
respectively.

\begin{figure}
\begin{center}
\vspace*{1em}
\includegraphics[width=7.5cm]{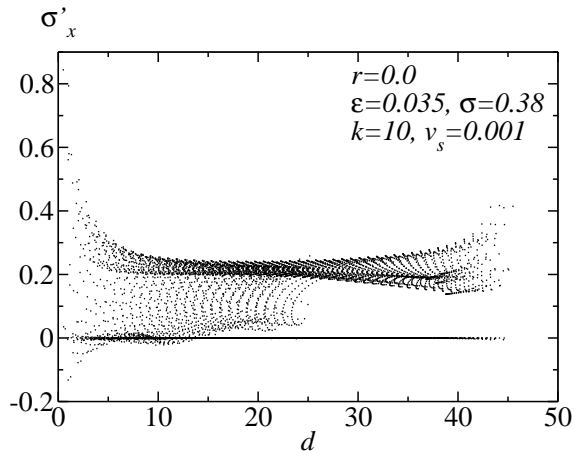}
\caption{\label{rs_r0.0}Spatial distribution of the $x$ components of 
spring forces, $\sigma'_{x}$, in the uniform system with $r=0$ 
when strain $\varepsilon=0.035$ and stress $\sigma=0.38$. }
\end{center}
\end{figure}

\begin{figure}
\begin{center}
\vspace*{1em}
\includegraphics[width=7.5cm]{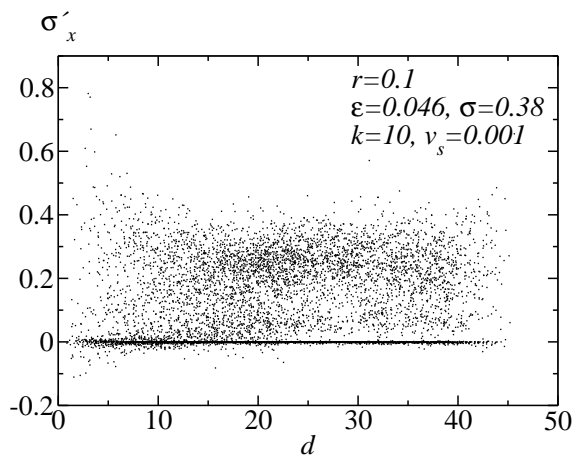}
\caption{\label{rs_r0.1}Same as Fig.~\ref{rs_r0.0} except in a sample of 
the disordered system with $r=0.1$
when strain $\varepsilon=0.046$ and stress 
$\sigma=0.38$.}
\end{center}
\end{figure}

\begin{figure}
\begin{center}
\vspace*{1em}
\includegraphics[width=7.5cm]{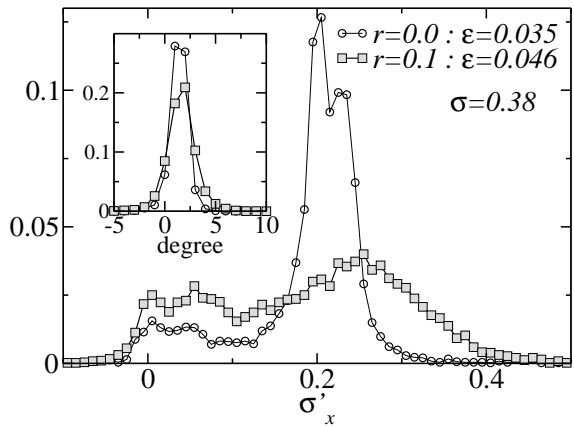}
\caption{\label{direc_Dx} 
Distribution of the force $\sigma'$ for the springs initially inclined $\pm 30^\circ$ 
to the $x$-axis.
Inset:  Distribution of angle displacements to the horizontal direction for the springs.
The stress value $\sigma=0.38$ is the same as Figs.~\ref{rs_r0.0} and \ref{rs_r0.1}. 
}
\end{center}
\end{figure}

We consider that this increase of the maximum stress should be connected
with differences between the stress distributions in the disordered systems
and that in the uniform systems.  
Figs.~\ref{rs_r0.0} and \ref{rs_r0.1} illustrate the $x$
component of each spring force $\sigma'_x$ against the distance between
the spring and the crack tip in the uniform system (Fig.~\ref{rs_r0.0})
and in the disordered system (Fig.~\ref{rs_r0.1}).
 The value of $\sigma$ is the same in the both figures. 
In the former case, 
 the forces are large only near the crack tip, and for $d>10$ most values
of the forces are distributed around $0.3$ or $0$.
Contrastingly, in the latter
case, there are many springs with forces beyond $0.3$ in the range
$d>10$. Thus, 
many springs distant from the crack tip suffer
larger stress in the disordered system than in the uniform system.   
It is more evident in Fig.~\ref{direc_Dx} which shows the 
distribution of the forces on the springs initially 
inclined  $\pm 30^\circ$ to the horizontal axis.
We find there that more springs share larger stress in the disordered 
system, which makes stress concentration at the crack tip to be eased.  

In our simulation, the lattice structure remains unchanged. We notice
concentration of data near the horizontal axis in Figs.~\ref{rs_r0.0}
and \ref{rs_r0.1}. They represent that the initially vertical springs do
not change their orientations and do not contribute to the stress, which
is true even near the crack tip. The other springs also keep their
orientations. The inset in Fig.~\ref{direc_Dx} shows the distribution of
angle displacements to the horizontal direction for the springs
initially $\pm 30^\circ$ inclined to the $x$ axis. The distribution is a
little moved to the horizontal direction and widened, but not so
much. This means that the lattice is only slightly deformed as the
system is stretched. This is due to the relatively small value of 
the threshold $\sigma^*=1$. Because of the smallness of the parameter, 
the fracture starts to expand at small strain $\varepsilon\cong 0.05$ for
$r=0.1$. At this value of strain, deformation of the lattice is very
small.

\begin{figure}
\begin{center}
\vspace*{1em}
\includegraphics[width=7.5cm]{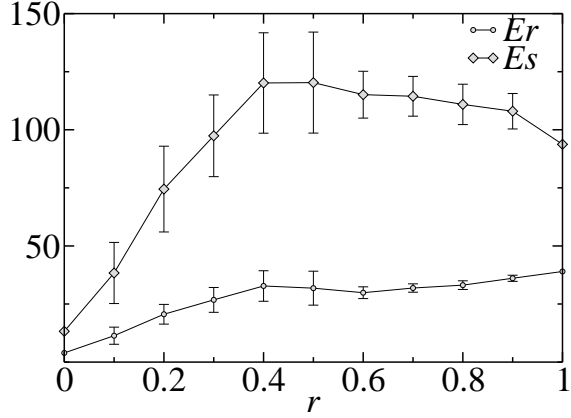}
\caption{\label{area} The injected energy $E_s$ and the released
 energy $E_r$ for several $r$}
\end{center}
\end{figure}

Not only the maximum stress but also the toughness of the system improves 
in the disordered system. The toughness is defined as the work required
to make a unit area of fracture surface in quasi-static fracture. 
We identify the work with the injected energy to the system
during advancing of cracks, $E_s$, which is defined by 
\begin{eqnarray}
E_s = H_0 W_0 A \varepsilon ,
\end{eqnarray}
where $A$ represents the area under the S-S curve. Though the estimate
contains other dissipative effects, they are expected to be small. We
assume further that the crack lengths do not vary with $r$. The upper
line in Fig.~\ref{area} shows $E_s$ averaged over $30$ samples, which
indicates that the toughness drastically increases with $r$ in region
$0<r<0.4$. In addition, the toughness in $0.4<r<1$ is larger than those
of the uniform systems ($r=0$ or $r=1$). 

\begin{figure}
\begin{center}
\vspace*{1em}
\includegraphics[width=7.5cm]{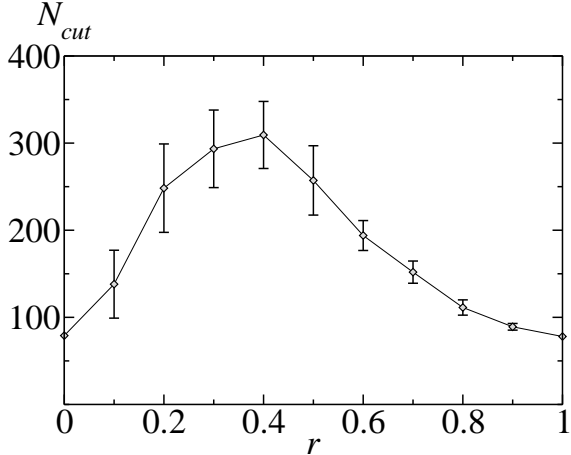}
\caption{\label{Ncut} Number of cut springs for several $r$.}
\end{center}
\end{figure}

\begin{figure}
\begin{center}
\vspace*{1em}
\includegraphics[width=7.5cm]{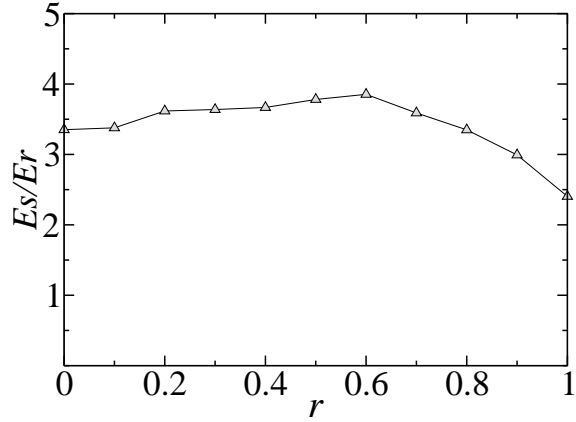}
\caption{\label{ErEs} $Es/E_r$ vs $r$.}
\end{center}
\end{figure}

A part of the injected energy is released by the fracture. The released
energy $E_r$ also changes with $r$ as shown in Fig.~\ref{area}. It 
increases with the ratio of soft springs $r$ except around $r=0.5$. This
is partly because a soft spring releases larger energy than a hard one
in our model; The spring with spring constant $k$ releases energy
${\sigma^*}^2/2k$. In addition, increase in the number of cut springs
also raises $E_r$. A hard spring is easily cut by smaller strain than a
soft spring. Thus, the crack tip proceeds with avoiding soft springs. As
the result, the crack meanders in the disordered systems, while it goes
straight in the uniform systems. The meandering causes the increase in
the number of cut springs, $N_{cut}$. We numerically observe that
$N_{cut}$ has the maximum near $r=0.4$ as shown in Fig.~\ref{Ncut}. The
inflection of $E_r$ around $r=0.5$ is due to the competition of the two
factors: the increase of soft springs which pushes up $E_r$ and the
decrease of $N_{cut}$. Precise relation between $E_s$ and $E_r$ is
unknown, but they largely exhibit similar change with
$r$. Fig.~\ref{ErEs} shows that the ratio $E_s/E_r$ does not change with
$r$ very much. Thus, the increase of $E_r$ in the disordered systems also
brings the increase of $E_s$, namely the increase of toughness.

\section{Conclusion and Discussion}

Through investigations of fracture in the disordered lattice systems
with unequal spring constants, we have found that both of the maximum
stress and the 
toughness improve in the disordered systems.
Because a soft spring holds larger energy than a hard one, and because
presence of soft springs makes fracture meander, the released energy
$E_r$ becomes larger in the disordered systems than the uniform ones.
This leads to the improvement of the toughness.

\begin{figure}
\begin{center}
\vspace*{2em}
\includegraphics[width=7.5cm]{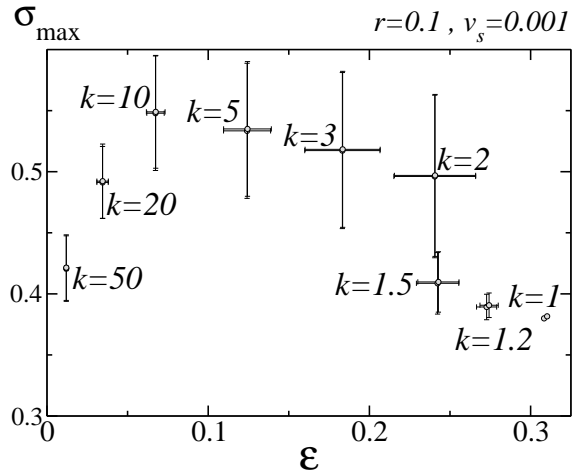}
\caption{\label{topSS} The maximum stress for several $k$ at $r=0.1$.
Each point is obtained from averaging over 30 samples.}
\end{center}
\end{figure}

Although we have so far fixed the ratio of the spring constants $k$ in the
simulations, improvement of maximum stress and fracture toughness  is
observed in a wide range of $k$ in our preliminary
computation. Fig.~\ref{topSS} shows the maximum stress
$\sigma_{\mathrm{max}}$ for disordered systems 
with various $k$ and fixed $r$ at $0.1$. Compared to the data for $r=0$
in Fig.~\ref{topSS_H}, we realize that the increase of the maximum stress 
is seen unless $k$ is too large. Moreover, $\sigma_{\mathrm{max}}$
increases with $k$ in the range $1\le k \le 10$ and decreases for 
$k>10$. For $k$ close to $1$, $\sigma_{\mathrm{max}}$ is small because
the system becomes nearly uniform. 
In contrast, if $k$ is too large, small strain causes large tension for 
hard springs but little effects on soft springs.
Therefore, most stress is sustained by hard springs only, and soft springs 
are not effective.
It means that the system behaves like diluted ones where
$\sigma_{\mathrm{max}}$ becomes small. We also observe that energy $E_s$
takes the maximum value at some $ 0 < r < 1$ if $k$ is distinctly larger
than $1$. Details of the computation will be reported elsewhere.

\begin{figure}
\begin{center}
\vspace*{2em}
\includegraphics[width=7.5cm]{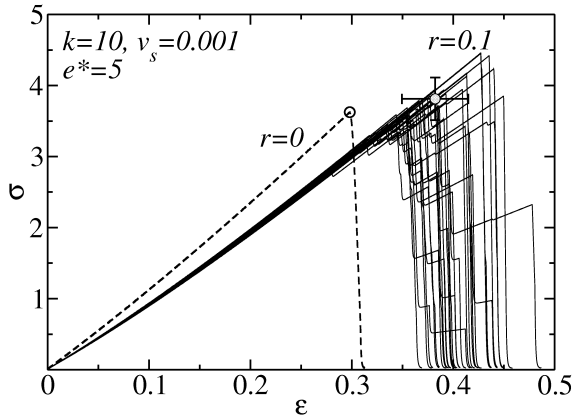}
\caption{\label{topSS_E} 
S-S curves for the uniform system ($r=0$) and the disordered system
 ($r=0.1, k=10$) where 
every spring breaks if its elastic energy exceeds a value $e^*$.
}
\end{center}
\end{figure}

Although the soft and hard springs are cut at the same stress value in our models,
this condition can be modified.  Let us consider the case where the springs are cut 
at the same elastic energy $e^*$.   Then, we again find that the toughness
improves in the disordered system for $r=0.1$, $k=10$, and $e^*=5$; 
For $r=0.0, 0.1$, the average $E_s=696$ and $941$, $E_r=390$ and $420$, respectively. 
Because the released energy when a spring is cut is common for all springs, 
$E_r$ is  proportional to the number of cut springs, $N_{cut}$.
Accordingly, the improvement of toughness in the disordered systems is caused as the
result of meandering crack.
The maximum stress also increases in some samples as shown in Fig.~\ref{topSS_E}.
However, their average represented by a circle point does not show a clear 
improvement.
We consider that the maximum stress should depend on $r$, and detailed study of
its dependence is planned for the future.

The fracture phenomenon naturally depends on the threshold stress $\sigma^*$.
It may drastically change with $\sigma^*$. Our choice that $\sigma^*=1$ is
rather small, and  fractures can expand with small strain below $0.3$.
Thus, we have observed fractures without large deformation of the
arrangement of the triangular lattice.  However, for materials like
polymer gels, fractures occur when the material suffers large
deformation. It is interesting to study if such large deformation can
improve the maximum stress or the toughness of the materials.

Our system is not large enough to be macroscopic.  We have checked that the system of
doubled size $101\times 100$ shows almost the same results.  However, it still may not 
be sufficient to restore isotropy.  Thus, the
orientation of the lattice may also have influences on the fracture. 
For example, if we use the system rotated by $90^\circ$, all spring forces have
components parallel to the external force.  Thus, the maximum stress can change
from our model where the vertical springs do not contribute to the
improvement of the toughness.
If we want to obtain results independent of orientation of the system,
use of random lattices will be effective because we can expect that 
isotropy is restored in systems as large as our model.  
It is a future problem to investigate such a system.

Our findings may be applied to design some new material with high
fracture toughness. However, our model is two dimensional and too simple
to apply real fractures. Some extensions, for example, to three
dimensions or to realistic interactions are future problems. 

\begin{acknowledgments}
One of the authors (C.U.) wishes to express gratitude to Akihiro
 Nakatani and So Kitsunezaki for stimulation and helpful discussions.
\end{acknowledgments}


\end{document}